\documentclass[fleqn]{llncs}

\usepackage{amsmath,amssymb}
\usepackage{hyperref}

\hypersetup{%
  pdftitle={Computing the Kullback-Leibler Divergence between two Generalized Gamma Distributions},
  pdfauthor={Christian Bauckhage},
  pdfsubject={data science, statistics, information theory},
  pdfkeywords={generalized Gamma distribution, Kullback-Leibler divergence},
  colorlinks=true,
  urlcolor=blue,
  citecolor=red,
  bookmarks=false}

\title{Computing the Kullback-Leibler Divergence between two Generalized Gamma Distributions}

\author{Christian Bauckhage}

\institute{%
  B-IT, University of Bonn, Bonn, Germany \\
  Fraunhofer IAIS, Sankt Augustin, Germany \\
  \email{http://mmprec.iais.fraunhofer.de/bauckhage.html}}

\begin{document}

\maketitle

\begin{abstract}
We derive a closed form solution for the Kullback-Leibler divergence between two generalized gamma distributions. These notes are meant as a reference and provide a guided tour towards a result of practical interest that is rarely explicated in the literature.
\end{abstract}

\section{The Generalized Gamma Distribution}

The origins of the \href{http://en.wikipedia.org/wiki/Generalized_gamma_distribution}{generalized gamma distribution} can be traced back to work of Amoroso in 1925 \cite{Amoroso1925-RIA,Crooks2010-TAD}. Here, we are concerned with the three-parameter version that was later introduced by Stacy \cite{Stacy1962-AGO}. Its probability density function is defined for $x \in [0,\infty)$ and given by
\begin{equation}
  \label{eq:GenGamma}
  f(x \mid a, d, p) = \frac{p}{a^d} \, \frac{x^{d-1}}{\Gamma \left(d/p\right)} \exp \left[- \left( \frac{x}{a} \right)^p \right]
\end{equation}
where $\Gamma(\cdot)$ is the \href{http://en.wikipedia.org/wiki/Gamma_function}{gamma function}, $a > 0$ determines scale and $d > 0$ and $p > 0$ are two shape parameters. We note that, depending on its parametrization, this unimodal density may be skewed to the left or to the right. Moreover, the generalized gamma contains other distributions as special cases. For $d = p$, it coincides with the Weibull distribution, and, if $p = 1$, it becomes the gamma distribution. Setting $d = p = 1$ yields the exponential distribution, and, for $a =2$, $p = 1$, and $d = k/2$ where $ k \in \mathbb{N}$, we obtain the $\chi^2$ distribution with $k$ degrees of freedom.

As a flexible skewed distribution, the generalized gamma is frequently used for life-time analysis and reliability testing. In addition, it models fading phenomena in wireless communication, has been applied in automatic image retrieval and analysis \cite{Choi2010-SWS,deVes2010-WBT,Schutz2013-CBT}, was used to evaluate dimensionality reduction techniques \cite{Li2006-IRP}, and also appears to be connected to diffusion processes in (social) networks \cite{Lienhard1967-APB,Bauckhage2013-TWA,Bauckhage2013-MMO}. Accordingly, methods for measuring (dis)similarities between generalized gamma distributions are of practical interest in data science because they facilitate model selection and statistical inference.

\section{The Kullback-Leibler Divergence}

The \href{http://en.wikipedia.org/wiki/Kullback–Leibler_divergence}{Kullback-Leibler divergence} (KL divergence) provides an asymmetric measure of the similarity of two probability distributions $P$ and $Q$ \cite{Kullback1951-OIA}. For the case where the two distributions are continuous, it is defined as
\begin{equation}
  \label{eq:DKL}
  D_{KL} (P \parallel Q) = \int\limits_{-\infty}^{\infty} p(x) \, \ln \frac{p(x)}{q(x)} \, dx
\end{equation}
where $p(x)$ and $q(x)$ denote the corresponding probability densities.

The KL divergence gives a measure of relative entropy. That is, it can be understood as the loss of information if $P$ is modeled in terms of $Q$. Hence, the smaller $D_{KL} (P \parallel Q)$, the more similar are $P$ and $Q$. Although this resembles the behavior of a distance measure, it is important to point out that the KL divergence does not define a distance since it is neither symmetric nor satisfies the triangle inequality.

\section{The KL Divergence between two Generalized Gamma Distributions}

Plugging two generalized gamma distributions $F_1$ and $F_2$ into \eqref{eq:DKL} and recalling that their probability densities $f_1$ and $f_2$ are defined for $x \in [0, \infty)$ yields
\begin{equation}
  \label{eq:DKLGG}
  D_{KL} (F_1 \parallel F_2) = \int\limits_{0}^{\infty} f_1(x \mid a_1, d_1, p_1) \ln \frac{f_1(x \mid a_1, d_1, p_1)}{f_2(x \mid a_2, d_2, p_2)} dx.
\end{equation}

\subsection{Step by Step Solution}

We begin evaluating the KL divergence in \eqref{eq:DKLGG} by considering the logarithmic factor inside the integral. For two generalized gamma densities as in \eqref{eq:GenGamma}, it is fairly easy to see that, after a few algebraic manipulations, this factor amounts to 
\begin{equation*}
\ln \frac{\frac{p_1}{a^{d_1}}}{\Gamma \left(\frac{d_1}{p_1}\right)} + (d_1-1) \ln x - \left( \frac{x}{a_1} \right)^{p_1} 
    - \ln \frac{\frac{p_2}{a^{d_2}}}{\Gamma \left(\frac{d_2}{p_2}\right)} - (d_2-1) \ln x + \left( \frac{x}{a_2} \right)^{p_2}
\end{equation*}
or, equivalently
\begin{equation}
\label{eq:log}
\underbrace{\ln \frac{\frac{p_1}{a^{d_1}}}{\Gamma \left(\frac{d_1}{p_1}\right)}
    - \ln \frac{\frac{p_2}{a^{d_2}}}{\Gamma \left(\frac{d_2}{p_2}\right)}}_{A} 
    + (d_1 - d_2) \ln x
    + \left( \frac{x}{a_2} \right)^{p_2} - \left( \frac{x}{a_1} \right)^{p_1}
\end{equation}
and we observe that term $A$ in \eqref{eq:log} is a constant independent of the variable of integration $x$.

Plugging \eqref{eq:log} back into \eqref{eq:DKLGG} then leads to 
\begin{align}
  & \int\limits_{0}^{\infty} f_1(x \mid a_1, d_1, p_1) \left[ A + (d_1 - d_2) \ln x + \left( \frac{x}{a_2} \right)^{p_2} - \left( \frac{x}{a_1} \right)^{p_1} \right] dx \notag \\
  = & \; A \, \int\limits_{0}^{\infty} f_1(x \mid a_1, d_1, p_1) \, dx \label{eq:IntConst}\\
  & +  \int\limits_{0}^{\infty} f_1(x \mid a_1, d_1, p_1) \, (d_1 - d_2) \ln x \, dx \label{eq:IntLog}\\
  & +  \int\limits_{0}^{\infty} f_1(x \mid a_1, d_1, p_1) \, \left( \frac{x}{a_2} \right)^{p_2} \, dx \label{eq:Intp2}\\
  & -  \int\limits_{0}^{\infty} f_1(x \mid a_1, d_1, p_1) \, \left( \frac{x}{a_1} \right)^{p_1} \, dx \label{eq:Intp1}
\end{align}

Given this expansion of \eqref{eq:DKLGG}, we consider the integrals in \eqref{eq:IntConst} to \eqref{eq:Intp1} one by one and then construct the final result from the intermediate results we thus obtain.

\subsubsection{Solving \eqref{eq:IntConst}} Since $f_1(x \mid a_1, d_1, p_1)$ is a probability density over $[0, \infty)$, we immediately see that 
\begin{equation}
\label{eq:IntConstFinal}
A \, \int_{0}^{\infty} f_1(x \mid a_1, d_1, p_1) \, dx = A \cdot 1 = A.
\end{equation}

\subsubsection{Solving \eqref{eq:IntLog}} Plugging the definition of $f_1(x \mid a_1, d_1, p_1)$ into \eqref{eq:IntLog} yields
\begin{align}
\label{eq:IntLogIntermediate}
& \int_{0}^{\infty} \frac{p_1}{a_1^{d_1}} \, \frac{x^{d_1-1}}{\Gamma \left(d_1/p_1\right)} \, e^{- \left( x / a_1 \right)^{p_1}} \, (d_1 - d_2) \, \ln x \, dx \notag \\[1ex]
= & \underbrace{\frac{p_1}{a_1^{d_1}} \, \frac{1}{\Gamma \left(\frac{d_1}{p_1}\right)} \, (d_1 - d_2)}_{B} \int_{0}^{\infty} x^{d_1-1} \, e^{- \left( x / a_1 \right)^{p_1}} \, \ln x \, dx
\end{align}
where we recognize the factor $B$ as another constant independent of $x$. In order to solve the integral in \eqref{eq:IntLogIntermediate}, we consider the following substitution
\begin{equation}
  \label{eq:subst}
  y = \left( \frac{x}{a_1}\right)^{p_1}
\end{equation}  
which is equivalent to 
\begin{equation}
y^{\frac{1}{p_1}} = \frac{x}{a_1}.
\end{equation}

Based on this substitution, we obtain new expressions for two of the factors inside the integral in \eqref{eq:IntLogIntermediate}, namely
\begin{align}
x^{d_1 - 1} & = a_1^{d_1 - 1} \, y^{\frac{d_1 -1}{p_1}} \\
\ln x & = \ln a_1 + \frac{1}{p_1} \, \ln y.
\end{align}
In addition, our substitution allows for rewriting the differential $dx$. In particular, we have
\begin{align}
dy & = \frac{p_1}{a_1} \, \left( \frac{x}{a_1} \right)^{p_1 - 1} \, dx \\
   & = \frac{p_1}{a_1} \, y^{\frac{p_1 - 1}{p_1}} \, dx
\end{align}
which is to say that
\begin{equation}
dx = \frac{a_1}{p_1} \, y^{- \frac{p_1 - 1}{p_1}} \, dy.
\end{equation}
Making use of all the above identities, the expression in \eqref{eq:IntLogIntermediate} can be recast and expanded as follows
\begin{align}
 & \; B \, \int_{0}^{\infty} a_1^{d_1 - 1} \, y^{\frac{d_1 -1}{p_1}} \, e^{-y} \, \ln a_1 \, \frac{a_1}{p_1} \, y^{- \frac{p_1 - 1}{p_1}} \, dy 
 \notag \\
 & \; + B \, \int_{0}^{\infty} a_1^{d_1 - 1} \, y^{\frac{d_1 -1}{p_1}} \, e^{-y} \, \frac{1}{p_1} \, \ln y \, \frac{a_1}{p_1} \, y^{- \frac{p_1 - 1}{p_1}} \, dy 
 \notag \\[2ex]
\label{eq:IntLogParts}  
= & \; B \, \frac{a_1^{d_1}}{p_1} \, \ln a_1 \, \int_{0}^{\infty} y^{\frac{d_1}{p_1}-1} \, e^{-y} \, dy + B \, \frac{a_1^{d_1}}{p_1^2} \, \int_{0}^{\infty} y^{\frac{d_1}{p_1}-1} \, e^{-y} \, \ln y \, dy
\end{align}

As the integrals in \eqref{eq:IntLogParts} are rather intricate, we next resort to the venerable text by Gradshteyn and Ryzhik \cite{Gradshteyn2007-TOI} which provides an invaluable resource for tackling integral equations. In particular, in \cite[eq. 3.381]{Gradshteyn2007-TOI}, we find
\begin{equation*}
\int_{0}^{\infty} y^{\nu-1} \, e^{-\mu y} \, dy = \frac{1}{\mu^\nu} \, \Gamma(\nu).
\end{equation*}
and \cite[eq. 4.352]{Gradshteyn2007-TOI} states that
\begin{equation*}
\int_{0}^{\infty} y^{\nu-1} \, e^{-\mu y} \, \ln y \, dy = \frac{1}{\mu^\nu} \, \Gamma(\nu) \, \bigl( \psi(\nu) - \ln \mu \bigr)
\end{equation*}
where $\psi(\cdot)$ denotes the \href{http://en.wikipedia.org/wiki/Digamma_function}{digamma function} for which we recall that 
\begin{equation*}
\psi(x) = \frac{d}{dx} \ln \Gamma(x) = \frac{\Gamma'(x)}{\Gamma(x)}.
\end{equation*}

Hence, if we set $\mu = 1$ and $\nu = d_1 / p_1$ and write out constant $B$ from \eqref{eq:IntLogIntermediate}, the first term in \eqref{eq:IntLogParts} becomes 
\begin{equation}
\frac{p_1}{a_1^{d_1}} \, \frac{1}{\Gamma \left(\frac{d_1}{p_1}\right)} \, (d_1 - d_2) \, \frac{a_1^{d_1}}{p_1} \, \ln a_1 \, \Gamma \left(\frac{d_1}{p_1}\right)
= (d_1 - d_2) \, \ln a_1
\end{equation} 
and for the second term in \eqref{eq:IntLogParts} we find 
\begin{equation}
\frac{p_1}{a_1^{d_1}} \, \frac{1}{\Gamma \left(\frac{d_1}{p_1}\right)} \, (d_1 - d_2) \, \frac{a_1^{d_1}}{p_1^2} \, \Gamma \left(\frac{d_1}{p_1}\right) \, \psi\left( \frac{d_1}{p_1} \right)  
= \frac{1}{p_1} \, \psi\left( \frac{d_1}{p_1} \right) \, (d_1 - d_2).
\end{equation}
Added back together, both expressions therefore provide us with the following, pleasantly simple intermediate result 
\begin{equation}
\label{eq:IntLogFinal}
\int\limits_{0}^{\infty} f_1(x \mid a_1, d_1, p_1) \, (d_1 - d_2) \ln x \, dx 
= \left[ \frac{1}{p_1} \, \psi\left( \frac{d_1}{p_1} \right) + \ln a_1 \right] \, (d_1 - d_2).
\end{equation}

\subsubsection{Solving \eqref{eq:Intp2}} Plugging the definition of $f_1(x \mid a_1, d_1, p_1)$ into \eqref{eq:Intp2} yields
\begin{equation}
\underbrace{\frac{p_1}{a^{d_1}} \, \frac{1}{\Gamma \left(\frac{d_1}{p_1}\right)}}_{C} \, \int_{0}^{\infty} x^{d_1-1} \, e^{- \left( x / a_1 \right)^{p_1}} \, \left( \frac{x}{a_2} \right)^{p_2} \, dx.
\end{equation}
In order to simplify this expression, we once again apply the substitution that was introduced in \eqref{eq:subst} and subsequently find 
\begin{align}
  & C \, \int_{0}^{\infty} a_1^{d_1 -1} \; y^{\frac{d_1-1}{p_1}} \, e^{-y} \, \left( \frac{a_1}{a_2} \right)^{p_2} \, y^{\frac{p_2}{p_1}} \, \frac{a_1}{p_1} \, y^{-\frac{p_1-1}{p_1}} \, dx \notag \\[2ex]
= \; & C \, \frac{a_1^{d_1}}{p_1} \, \left( \frac{a_1}{a_2} \right)^{p_2} \, \int_{0}^{\infty} y^{\frac{d_1 + p_2}{p_1} - 1} \; e^{-y} \, dy.
\end{align}

Looking at this integral, we recognize its structure to be similar to that of the integral in the first term of \eqref{eq:IntLogParts}. As we already know how to deal with integrals like these, we omit further details and immediately obtain our next intermediate result
\begin{equation}
\label{eq:Intp2Final}
\int\limits_{0}^{\infty} f_1(x \mid a_1, d_1, p_1) \,  \left( \frac{x}{a_2} \right)^{p_2} \, dx 
= \frac{\Gamma \left(\frac{d_1+p_2}{p_1}\right)}{\Gamma \left(\frac{d_1}{p_1}\right)} \, \left( \frac{a_1}{a_2} \right)^{p_2}.
\end{equation}

\subsubsection{Solving \eqref{eq:Intp1}} Finally, plugging the definition of $f_1(x \mid a_1, d_1, p_1)$ into \eqref{eq:Intp1} yields
\begin{equation}
\label{eq:Intp1Intermediate}
-C \, \int_{0}^{\infty} x^{d_1-1} \, e^{- \left( x / a_1 \right)^{p_1}} \, \left( \frac{x}{a_1} \right)^{p_1} \, dx.
\end{equation}
where the multiplicative constant $C$ is defined as above. Again, this expression can be solved quickly using the change of variables we considered  before. That is, applying \eqref{eq:subst}, the expression in \eqref{eq:Intp1Intermediate} can be written as
\begin{align}
  & -C \, \int_{0}^{\infty} a_1^{d_1 -1} \; y^{\frac{d_1-1}{p_1}} \, e^{-y} \, y \, \frac{a_1}{p_1} \, y^{\frac{-p_1-1}{p_1}} \, dy \notag \\[2ex]
= \; & -C \, \frac{a_1^{d_1}}{p_1} \, \int_{0}^{\infty} y^{\frac{d_1}{p_1}} \, e^{-y} \, dy
\end{align}
where we have used that $y = y^{p_1 / p_1}$. Once more, we recognize a structural similarity to the first term in \eqref{eq:IntLogParts} so that, if we set $\nu = d_1 / p_1 + 1$, we obtain our final intermediate result as follows
\begin{equation}
\label{eq:Intp1Final}
- \int\limits_{0}^{\infty} f_1(x \mid a_1, d_1, p_1) \,  \left( \frac{x}{a_1} \right)^{p_1} \, dx 
= - \frac{\Gamma \left(\frac{d_1}{p_1} + 1\right)}{\Gamma \left(\frac{d_1}{p_1}\right)} = - \frac{d_1}{p_1}.
\end{equation}

\subsection{Final Result}

Finally, assembling the four intermediate results in \eqref{eq:IntConstFinal}, \eqref{eq:IntLogFinal}, \eqref{eq:Intp2Final}, and \eqref{eq:Intp1Final} establishes that: The KL divergence between two generalized gamma densities $f_1$ and $f_2$ amounts to
\begin{align}
\label{eq:final}
  & \int\limits_{0}^{\infty} f_1(x \mid a_1, d_1, p_1) \ln \frac{f_1(x \mid a_1, d_1, p_1)}{f_2(x \mid a_2, d_2, p_2)} \notag \\
= & \ln \frac{p_1 \, a_2^{d_2} \, \Gamma\left(\frac{d_2}{p_2}\right)}{p_2 \, a_1^{d_1} \, \Gamma\left(\frac{d_1}{p_1}\right)} 
    + \left[ \frac{\psi\left( \frac{d_1}{p_1} \right)}{p_1} + \ln a_1 \right]  (d_1 - d_2) 
    + \frac{\Gamma\left(\frac{d_1+p_2}{p_1}\right)}{\Gamma\left(\frac{d_1}{p_1}\right)} \left( \frac{a_1}{a_2} \right)^{p_2} 
    - \frac{d_1}{p_1}
\end{align}

\section{Concluding Remarks}

Given this closed form solution for the KL divergence between two generalized gamma distributions, it appears instructive to verify it for special cases. Equating the shape parameters $d$ and $p$ of a generalized gamma distribution produces a \href{http://en.wikipedia.org/wiki/Weibull_distribution}{Weibull distribution}. Evaluating \eqref{eq:final} for the particular case where $d_1 = p_1$ and $d_2 = p_2$ yields
\begin{equation}
\ln \frac{p_1 \, a_2^{p_2}}{p_2 \, a_1^{p_1}} 
+ \left[ - \frac{\gamma}{p_1} + \ln a_1 \right]  (p_1 - p_2) 
+ \Gamma\left( \frac{p_2}{p_1} + 1 \right) \left( \frac{a_1}{a_2} \right)^{p_2} 
- 1
\end{equation}
where $\gamma = -\psi(1) \approx  0.5772$ is the \href{http://en.wikipedia.org/wiki/Euler_constant}{Euler constant}. This indeed corresponds to the KL divergence between two Weibull distributions \cite{Bauckhage2013-CTK}.

Likewise, equating the shape parameter $p$ of a generalized gamma distribution to $1$ produces a \href{http://en.wikipedia.org/wiki/Gamma_distribution}{gamma distribution}. An evaluation of \eqref{eq:final} for the special case where $p_1 = p_2 = 1$ yields
\begin{equation}
\ln \frac{a_2^{d_2} \, \Gamma (d_2)}{a_1^{d_1} \, \Gamma(d_1)} 
    + \left[ \psi(d_1) + \ln a_1 \right]  (d_1 - d_2) 
    + d_1 \left( \frac{a_1}{a_2} \right)
    - d_1
\end{equation}
which is indeed the arguably well known KL divergence between two gamma distributions.

\bibliographystyle{splncs}
\bibliography{literature}

\begin{thebibliography}{10}

\bibitem{Amoroso1925-RIA}
Amoroso, L.:
\newblock {Ricerche intorno alla curve dei redditi}.
\newblock Annali di Matematica Pura ed Applicata \textbf{2}(1) (1925)  123--159

\bibitem{Crooks2010-TAD}
Crooks, G.:
\newblock {The Amoroso Distribution}.
\newblock arXiv:1005.3274 [math.ST] (2010)

\bibitem{Stacy1962-AGO}
Stacy, E.:
\newblock {A Generalization of the Gamma Distribution}.
\newblock The Annals of Mathematical Statistics \textbf{33}(3) (1962)
  1187--1192

\bibitem{Choi2010-SWS}
Choi, S., Tong, C.:
\newblock {Statistical Wavelet Subband Characterization based on Generalized
  Gamma Density and Its Application to Texture Retrieval}.
\newblock IEEE Trans. on Image Processing \textbf{19}(2) (2010)  281--289

\bibitem{deVes2010-WBT}
{de Ves}, E., Benavent, X., Ruedin, A., Acevedo, D., Seijas, L.:
\newblock {Wavelet-based Texture Retrieval Modeling the Magnitudes of Wavelet
  Detail Coefficients with a Generalized Gamma Distribution}.
\newblock In: Proc. ICPR. (2010)

\bibitem{Schutz2013-CBT}
Schutz, A., Bombrum, L., Berthoumieu, Y., Najim, M.:
\newblock {Centroid-Based Texture Classification Using the Generalized Gamma
  Distribution }.
\newblock In: Proc. EUSIPCO. (2013)

\bibitem{Li2006-IRP}
Li, P., Hastie, T., Church, K.:
\newblock {Improving Random Projections Using Marginal Information}.
\newblock In: Proc. COLT. (2006)

\bibitem{Lienhard1967-APB}
Lienhard, J., Meyer, P.:
\newblock {A Physical Basis for the Generalized Gamma Distribution}.
\newblock Quarterly of Applied Mathematics \textbf{25}(3) (1967)  330--334

\bibitem{Bauckhage2013-TWA}
Bauckhage, C., Kersting, K., Rastegarpanah, B.:
\newblock {The Weibull as a Model of Shortest Path Distributions in Random
  Networks}.
\newblock In: Proc. Int.~Workshop on Mining and Learning with Graphs, Chicago,
  IL, USA, ACM (2013)

\bibitem{Bauckhage2013-MMO}
Bauckhage, C., Kersting, K., Hadiji, F.:
\newblock {Mathematical Models of Fads Explain the Temporal Dynamics of
  Internet Memes}.
\newblock In: Proc. ICWSM, AAAI (2013)

\bibitem{Kullback1951-OIA}
Kullback, S., Leibler, R.:
\newblock {On Information and Sufficiency}.
\newblock Annals of Mathematical Statistics \textbf{22}(1) (1951)  79--86

\bibitem{Gradshteyn2007-TOI}
Gradshteyn, I., Ryzhik, I.:
\newblock {Tables of Integrals, Series, and Products}. 7th edn.
\newblock Academic Press (2007)

\bibitem{Bauckhage2013-CTK}
Bauckhage, C.:
\newblock {Computing the Kullback-Leibler Divergence between two Weibull
  Distributions}.
\newblock arXiv:1310.3713 [cs.IT] (2013)

\end{thebibliography}

\end{document}